\documentclass[12pt,preprint]{aastex61}
\usepackage{color}
\newcommand{\be}{\begin{equation}}
\newcommand{\ee}{\end{equation}}
\def\ergs{{\rm\,erg\,s^{-1}}}
\newcommand{\msun}{{M}_{\sun}}

\shorttitle{Location of $\gamma$-ray emitting region in blazars} \shortauthors{Lin et al. }
\begin{document}

\title{Constraints on the location of $\gamma$-ray emitting region for a sample of blazars with radio core-shift measurements}

\correspondingauthor{Qingwen Wu}
\email{qwwu@hust.edu.cn}

\author{Linhui Wu}  
\affiliation{School of Physics, Huazhong University of Science and Technology, Wuhan 430074, China}
 
 \author{Qingwen Wu} 
\affiliation{School of Physics, Huazhong University of Science and Technology, Wuhan 430074, China}

\author{Dahai Yan} 
\affiliation{Yunnan Observatory, Chinese Academy of Sciences, Kunming 650011, China}

\author{Liang Chen} 
\affiliation{Shanghai Astronomical Observatory, Chinese Academy of Science, 80 Nandan, Shanghai 200030, China}
\affiliation{University of Chinese Academy of Science, 19A Yuquanlu, Beijing 100049, China}

\author{Xuliang Fan} 
\affiliation{School of Physics, Huazhong University of Science and Technology, Wuhan 430074, China}

\begin{abstract}
  We model simultaneous or quasi-simultaneous multi-band spectral energy distributions (SEDs) for a sample of 25 blazars that have radio core-shift measurements, where a one-zone leptonic model and Markov chain Monte Carlo technique are adopted. In the SED fitting for 23 low-synchrotron-peaked (LSP) blazars, the seed photons from broad line region (BLR) and molecular torus are considered respectively in the external Compton process. We find that the SED fitting with the seed photons from the torus are better than those utilizing BLR photons, which suggests that the $\gamma$-ray emitting region may be located outside the BLR. Assuming the magnetic field strength in the $\gamma$-ray emitting region as constrained from the SED fitting follows the magnetic field distribution as derived from the radio core-shift measurements (i.e., $B(R)\simeq B_{\rm 1 pc}(R/\rm 1pc)^{-1}$, $R$ is the distance from the central engine and $B_{\rm 1 pc}$ is the magnetic field strength at 1 pc), we further calculate the location of the $\gamma$-ray emitting region, $R_{\gamma}$, for these blazars. We find that $R_{\gamma}\sim2\times10^{4}R_{\rm S}\simeq10 R_{\rm BLR}$ ($R_{\rm S}$ is the Schwarzschild radius and $R_{\rm BLR}$ is the BLR size), where $R_{\rm BLR}$ is estimated from the broad line luminosities using the empirical correlations obtained using the reverberation mapping methods.
\end{abstract}


\section{Introduction}
Blazars, further sub-classified into BL Lacertae objects (BL Lacs) and flat-spectrum radio quasars (FSRQs), are a special subclass of active galactic nuclei (AGNs) with relativistic jets moving close to our line of sight, where the FSRQs normally have strong emission lines while BL Lacs have weak or no emission lines \cite[e.g.,][]{urry95}. Due to strong boosting effect, blazars normally produce fully nonthermal strongly polarized radiation across the electro-magnetic spectrum from radio to $\gamma$-rays, they are extremely luminous, and can exhibit superluminal motion effect, among other features \cite[e.g.,][]{brin86,bege87,ulri97,jors01,fan12,fan15,fan16,fan17}. There are two most evident peaks in the multi-wavelength spectral energy distribution (SED) of blazars where, in the leptonic model, the lower energy peak is believed to be dominated by synchrotron radiation while the radiation at the higher energy peak could be due to the inverse Compton scattering \cite[e.g.,][]{bott07}. Based on the frequency of the first peak, blazars are further divided into low-synchrotron-peaked (LSP; i.e.,$\nu_{p}^{S}<10^{14}$ Hz), intermediate-synchrotron-peaked (ISP; i.e.,$10^{14}<\nu_{p}^{S}<10^{15}$ Hz), and high-synchrotron-peaked (HSP; i.e.,$\nu_{p}^{S}>10^{15}$ Hz) blazars \cite[e.g.,][]{pado95,abdo10a}. The Fermi Large Area Telescope (LAT) was launched in 2008 and has detected serveral thousand blazars in the $\gamma$-ray energy region (e.g., more than 3000 sources in 3rd Fermi Large Area Telescope source catalog, \citealt{acer15}; see also \citealt{abdo10b,nola12,acer15,acke15}), which give a good opportunity to explore the physical properties of their central engines.

The multi-wavelength emission in blazars plays an important role in helping us to understand their central engines and particularly in jet physics (e.g., formation, acceleration, collimation, radiation and composition). Submillimeter observations are now starting to reach the angular resolution sufficient to resolve the region of the BH horizon and jet-formation region in nearby supermassive black hole systems \cite[e.g., Sgr A* and M 87,][]{doel12}. However, the angular resolution is still quite limited for $\gamma$-ray telescopes and it will not be nearly as high as in submillimeter waveband in any foreseeable future. Therefore, one cannot resolve the site of the $\gamma$-ray emitting region in blazars now and even in near future, which prevents us from learning the possible physics of the acceleration and dissipation of the jet further. 

Based on indirect evidence, there are two main candidates for the $\gamma$-ray emitting region. The first one is close to the BH and the $\gamma$-ray emission is produced inside the broad line region \cite[BLR, e.g.,][]{derm93,blan95,ghis96,geor01,fan06,bai09,tave09,isle13,hu15}. This argument is supported by constraining the size of the $\gamma$-ray emission region through the magnification factor during the $\gamma$-ray variability in some gravitationally lensed $\gamma$-ray blazars \cite[e.g.,][]{nero15,vovk16}, the short variability time scales \cite[down to several hours, e.g.,][]{tave10} and the sharp breaks at GeV band seen in the $\gamma$-ray spectra of some FSRQs that may caused by the opacity to pair production \cite[e.g.,][]{liu06,bai09,pout10,ster11}. The second possible region for the $\gamma$-rays production is beyond the BLR \cite[e.g.,][]{blaz00,arbe02,soko05,yan12,meye15}. Some multi-wavelength observations for the giant flares suggested that both the sub-millimeter and $\gamma$-rays are produced 10-20 pc from the BH \cite[e.g.,][]{lari08,siko08}, where it was assumed that gamma-ray and radio emission is triggered by shocks propagating along a relativistic jet. The time delay between flares in radio and gamma-ray bands combined with the VLBI high-resolution observation on the size of radio core also indicated that the $\gamma$-ray emitting region stays in the upstream of, but not far from the radio core \cite[e.g.,][]{push10,mars10,jors10,agud11a,fuhr14,max14,rama16}. It is normally believed that $\gamma$-ray emission in blazars come from inverse-Compton (IC) process, where external Compton (EC) plays an important role in LSP blazars while ISP and HSP blazars ( BL Lacs) generally can be explained by the synchrotron self-Compton (SSC) \cite[e.g.,][]{kraw04,chen11,zhang14,kang14}. In the EC process, the seed photons are determined by the location of $\gamma$-ray emitting region, which may be dominated by accretion disk, BLR, infrared torus and cosmic background respectively if the gamma-ray emitting region is located near the BH horizon, inside the BLR, inside the torus or much beyond of the torus \cite[e.g.,][]{ghis09}. \cite{kang14} compared the seed photons from the BLR and dusty torus in SED fitting of LSP blazars, and found that the seed photons from torus are preferred based on a $\chi^2$ method. \cite{zheng17} also explored this issue using a stratified jet and found that the seed photons from torus provide better match to observations.

 Opacity-driven shifts of the apparent radio core position with frequency, known as the core-shift effect, is another promising tool to explore the jet physics in radio AGNs. The VLBI technique provide the jet images of extragalactic SMBH systems with spatial resolution of an order of pc-scale, which is slightly larger but comparable the size of BLR in many nearby AGNs \cite[e.g.,][]{pear96,zens97,zens06,loba10}. The apparent outward shift of the VLBI core position with decreasing observation frequency can be attributed to the synchrotron self-absorption process in the jet \cite[e.g.,][]{blan79}. Investigation of the core-shift effect can provide deeper understanding of the structure and physical conditions in AGN jets \cite[e.g., core size, magnetic field strength,][]{loba98,hiro05}. The core position, $R_{\rm c}$, as a function of frequency, $\nu$, is found to roughly follow a $R_{\rm c} \propto \nu^{-1}$ law \cite[e.g.,][]{soko11,osul09}, which is consistent with the prediction for a synchrotron self-absorbed conical jet in equipartition \citep[e.g.,][]{koni81,loba98}.

 Even though there are fruitful $\gamma$-ray observations for blazars, the exact location of the $\gamma$-ray production site remains unclear. Using the measured core-shifts, the distribution of equipartition magnetic field strength along the core size can be derived at $\sim$ pc scale \cite[$B-R_{\rm c}$ relation, e.g.,][]{osul09}. In SED fitting of blazars, the magnetic field strength can also be estimated for the $\gamma$-ray emitting region \cite[e.g.,][]{zhang12,kang14,yan14,ghis15}. It should be pointed out that the positions of the pc-scale radio cores are close to the expected $\gamma$-ray emitting region. Therefore, it is possible to calculate the \textit{exact} location of the $\gamma$-ray emitting region assuming the magnetic field follows the $B-R_{\rm c}$ relation, which is the motivation of this work. In Section 2, we present our sample. The model and results are shown in Sections 3 and 4 respectively. The discussion and conclusion are presented in Section 5. Throughout this work, the base 10 logarithms are adopted, and we assume the following cosmology: $H_{0}=70\ \rm km\ s^{-1} Mpc^{-1}$, $\Omega_{M}=0.3$ and $\Omega_{\Lambda}=0.7$.

\section{The sample}
For the purpose of this work, we select our blazar sample using the following criteria: 1) the source should have a radio core-shift measurement, as selected from \cite{zama14} and \cite{osul09}; 2) there are simultaneous or quasi-simultaneous multi-wavelength observational data available, where we selected the data from \cite{abdo10a} and \cite{giom12} (more data points will be selected if the sources are included in both papers). In total, we selected 25 blazars (see Table 1) and the SEDs are presented in Figures 1-6, where Figure 1 particularly shows the SED fitting of BL Lac. There are 16 sources selected from \cite{giom12}, where the radio to X-ray data are observed simultaneously, while the $\gamma$-ray data have three types: simultaneous observations during the $Planck$ observation shown with red filled circles, quasi-simultaneous observations within the 2 months centered on the $Planck$ observation shown with green filled squares, and 27 months of Fermi-LAT integration shown with blue open triangles. The other 9 sources are selected from \cite{abdo10a}, where the $\gamma$-ray data observed between 2008 August and October with Fermi are presented with green filled squares, and other simultaneous and quasi-simultaneous multi-wavelength data are shown with the red filled circle and green filled square points respectively (mainly between 2008 May and 2009 January). The instruments contributing data to the sample are $Fermi$, $Swift$, $GASP$-$WEBT$, $KVA$, $Xinglong$, $Planck$, $APEX$, $ATCA$, $Effelsberg$, $IRAM$, $Medicina$, $Metsahovi$, $OVRO$, $RATAN$-$600$, $UMRAO$ and $VLA$ \cite[for details see][]{abdo10a,giom12}. The IAU name, redshift, source type and data selection for $\gamma$-ray emission are listed in Table (1).    

 \section{Model and Method}
We adopt a simple, one-zone, homogeneous leptonic synchrotron and IC scattering model that is widely used in fitting the SED of blazars \cite[][and references therein]{ghis09}, which can be found in \cite{kang14}, we simply describe the model here. The non-thermal radiation is produced in a spherical blob (radius of $r_{\rm blob}$) filled with a uniform magnetic field ($B$), which is moving relativistically at a small angle to our line of sight, where the Doppler factor $\delta = [\Gamma(1-\beta \cos\theta)]^{-1} \thickapprox \Gamma$. The size of the blob is estimated from the minimum variability timescale ($\Delta t_{\rm var}$) with $r_{\rm blob}=c\delta \Delta t_{\rm var}/(1+z)$, where $c$ is light speed, $z$ is redshift of the source. We use a broken power-law electron energy distribution,
 \be
  N(\gamma )=\left\{ \begin{array}{ll}
                    N_{0}\gamma ^{-p_1}  &  \mbox{ $\gamma_{\rm min}\leq \gamma \leq \gamma_{\rm b}$} \\
            N_{0}\gamma _{\rm b}^{p_2-p_1} \gamma ^{-p_2}  &  \mbox{ $\gamma _{\rm b}<\gamma\leq\gamma_{\rm max}$,}
           \end{array}
       \right.
   \label{Ngamma}
  \ee
where $N_{0}$ denotes the normalization of the electron distribution, $\gamma_{\rm b}$ denotes the break Lorentz factor, $p_{1}$ and $p_{2}$ denote the indices of the power law below and above $\gamma_{\rm b}$, $\gamma_{\rm min}$ and $\gamma_{\rm max}$ are the minimum and maximum electron Lorentz factor respectively.

To explain the higher-energy peak in SEDs of our blazars, we consider both SSC and EC in our SED fitting, where the EC component is normally not important in BL Lacs. The Klein-Nishina (KN) effect in the IC scattering \cite[see][]{blum70,rybi79} as well as the photon-photon absorption by extragalactic background light (EBL) are considered \cite[see][]{fink10}. For the EC process, we consider the seed photons either predominantly come from the BLR or the dusty torus, depending on the location of the $\gamma$-ray emitting region ($R_{\gamma}$). If $R_{\gamma}< R_{\rm BLR}$, we adopt a constant energy density of the BLR seed photons, $u_{\rm BLR}\simeq 2.65\times10^{-2}$erg cm$^{-3}$, where the seed photon energy density $u_{\rm BLR}\simeq f_{\rm BLR}L_{\rm d}/4\pi c R_{\rm BLR}^2$, $f_{\rm BLR}$=0.1, is the fraction of disk luminosity $L_{\rm d}$ that is re-emitted by the broad emission lines and $R_{\rm BLR}\simeq 10^{17}L_{\rm d,45}^{1/2}$ cm \cite[through the reverberation mapping technique, e.g.,][]{kasp07,bent09}. In the jet comoving frame, $u_{\rm BLR}^{'}\simeq (17/12)\Gamma^{2}u_{\rm BLR}$ \cite[see][for details]{ghis08,ghis09}. The radiation can be described by an isotropic blackbody of a peak frequency at $2\times10^{15}\Gamma$ Hz predominantly contributed by Ly$\alpha$ lines \cite[e.g.,][]{ghis08}. For the case of $R_{\gamma}>R_{\rm BLR}$, the seed photons from the dusty torus will dominate. Similar to $u_{\rm BLR}$, $u_{\rm torus}\simeq f_{\rm torus}L_{\rm d}/4\pi c R_{\rm torus}^2$, where $f_{\rm torus}\simeq0.5$ and $R_{\rm torus} \simeq 2.5\times10^{18}L_{\rm d,45}^{1/2}$ cm \cite[see][for more details]{ghis08}. The torus reprocesses the accretion-disk radiation into the infrared band with a typical temperature around 15-200 K \citep[e.g.,][]{clea07}. In this work, we adopt the energy density of soft photons in torus of $u_{\rm torus}^{'}\simeq 3\times10^{-4}\Gamma^{2}$ erg cm$^{-3}$ in the jet comoving frame and a blackbody radiation with a peak frequency at $\sim 3\times10^{13}\Gamma$ Hz \citep[e.g.,][]{ghis08}. It should be noted that the seed photon energy density of both BLR and torus are roughly constant in the lab frame if the $\gamma$-ray emitting region is smaller than the BLR or torus due to both $R_{\rm BLR}$ and $R_{\rm torus}$ being roughly proportional to $L_{\rm d}^{1/2}$ \cite[e.g.,][]{ghis09}.

In total, there are nine parameters in fitting of the SEDs: $r_{\rm blob}$, $B$, $\delta$, $p_{1}$, $p_{2}$, $\gamma_{min}$, $\gamma_{max}$, $\gamma_{b}$, and $N_{0}$, where $B$ is the magnetic field strength in the spherical blob. For the size of the blob, we estimate from the minimum variability timescale, $\Delta t_{\rm var}$, where $\Delta t_{\rm var}$ is selected from literature (nine sources shown in Table 1) or takes a typical value of 1 day if there are no observational constraints \cite[e.g.,][]{ghis98,abdo09,foss08,zhang12,cao13}. We set $\gamma_{\rm max} = 100\gamma_{\rm b}$ in this work,  because our model is not very sensitive to this parameter. Therefore, we have seven free parameters in our SED fittings. We adopt an MCMC method based on Bayesian statistics, in estimating the model parameters, superior to the grid approach in efficiency of sampling the parameter space \cite[also see][]{yan13,yan15,zhu16}. The jumping strategy in the parameter apace is determined by the Metropolis-Hastings sampling algorithm \citep{mack03}. The algorithm ensures the number density of samplings statistically approach the probability density functions of the model parameters. In the SED fittings, we run single chains with the Raftery $\&$ Lewis convergence diagnostics \citep[][]{raft92}, and flat priors in the model parameters are assumed. Introduction to MCMC sampling can be seen in \cite{neal93,game97,mack03,fan10}, and the code we used in this paper is obtained from  COSMOMC\footnote{\software{COSMOMC \citep{lewi02}}}, see \cite{lewi02} for more details. In the fitting, we adopt the relative systematic uncertainty of $5\%$ for the optical-UV data \cite[e.g.,][]{pool08}, $5\%$ for X-rays data reported in \cite{abdo11} and $10\%$ for $\gamma$-ray data \cite[e.g.,][]{acke12}. 

To compare the $\gamma$-ray emitting region with the BLR size, we estimate the BLR size from the empirical correlations derived from the reverbration mapping method \citep[e.g.,][]{kasp00}, where the correlations of $R_{\rm BLR}-L_{\rm H\beta}$, $R_{\rm BLR}-L_{\rm Mg II}$ and $R_{\rm BLR}-L_{\rm C IV}$ are adopted from \cite{wu04} and \cite{kong06} respectively (see Table 2 for the line types and line width). The virial BH mass of FSRQs are calculated from the line width and BLR size (equation 5 in \citealt{kasp00} for H$\beta$ emission line, equations 5 and 6 in \citealt{kong06} for C IV and Mg II lines).

\section{Results}
  Similarly to former works \citep[e.g.,][]{zhang10,zhang12,kang14,kang16,yan16}, we neglect the low-frequency radio data and consider the data with $\nu>200$ GHz (or $\log \nu>11.3$) in our SED fitting with the one-zone model due to the fact that radio emission should come from the large-scale jet and cannot be accounted for with a one-zone model. The variability correlation between millimeter, optical, X-ray and $\gamma$-ray emission support that they come from more or less similar region \cite[e.g.,][]{siko08,leon12,wehr12,damm13,orie13}. In 4 LSP blazars (0333+321, 0430+052, 2145+067, 2230+114), the putative UV excesses are not included in our SED fitting, as they should come from the cold accretion disk \cite[][]{shak73,ghis13,ajel16}. On average, there are 17 data points in our fitting. As an example, we show the multi-wavelength SED and its fitting for 2200+420 (BL Lac) in Figure 1 (left panel), where only the SSC process is considered due to the EC not being important in this source. In the right panel of Figure 1, we show the probability distributions of the model parameters, where $B = 0.3_{-0.1}^{+0.4}$ G, $\delta = 10.1_{-2.8}^{+2.9}$, $p_{1} = 1.9_{-0.2}^{+0.2}$, $p_{2}=4.2_{-0.2}^{+0.3}$, $\gamma_{\rm b}=37_{-12}^{+17}\times10^{2}$, $N_0=0.5_{-0.4}^{+0.6}\times10^{4}$, $\gamma_{\rm min}=5_{-4}^{+39}$ (upper and lower limits represent 1$\sigma$ errors, see also Table 1). The source of 1219+285 is also a BL Lac object, for which EC component is negligible. For the remaining 23 LSP sources, both SSC and EC component are considered, where the SEDs and the fitting are shown in Figures 2-6. For each source, the SED fitting with seed photons from torus and BLR are presented in left and right panels respectively. In our SED \text{fitting}, we find that most of LSP blazars have $\chi^{2}_{\rm BLR}/\chi^{2}_{\rm torus}>1$ (20 of 23 sources, $\chi^{2}_{\rm BLR}/\chi^{2}_{\rm torus}\sim1$ for two other sources and only one has the ratio $\sim 0.6$, see Table 1), where the distribution of the ratio $\chi^{2}_{\rm BLR}/\chi^{2}_{\rm torus}$ is shown in Figure 7. Our results suggest that the SED fitting with the seed photons from torus are better than with those from the BLR, which support the notion that the location of the $\gamma$-ray emitting region should stay outside of the BLR. In the following work, therefore, we consider the model parameters from the SED fittings with the torus seed photons.
  
 In our models, we find the magnetic field strength is normally around 1 Guass (average value $<B>=1.2$ G) in the blazars of our sample, where the $B$ is a little bit weaker in two BL Lac objects (2200+420 and 1219+285, $B\sim$0.1-0.3 G). The magnetic field strength in the $\gamma$-ray emitting region is more or less similar to that in pc-scale jet as estimated from the core-shift effect \cite[e.g.,][]{from13,zama14,agar17}. In the jet core-shift studies, it was found that $B\propto R^{\sim -1}$ within the region of radio cores \citep[e.g.,][]{osul09,soko11,from13,agar17}. Assuming the magnetic field strength in the $\gamma$-ray emitting region as constrained from blazar SED \text{fitting} roughly follows the magnetic field distribution as derived from the core-shift effect, using $B=B_{1}R^{-1}$ ($B_1$ is the magnetic field strength at 1 pc), we calculate position of the $\gamma$-ray emitting region $R_{\gamma}$ for each source in our sample. As an example, we show the magnetic field distribution and the location of the possible $\gamma$-ray emitting region of $R_{\gamma}$ in Figure 8. We find that the $\gamma$-ray emitting region is located $\sim 10^{3-5}R_{\rm S}$ from the central engine, with an average value of $\sim 2\times10^{4}R_{\rm S}$, where the distribution of $R_{\gamma}/R_{\rm S}$ are shown in Figure 9. We also present the distribution of $R_{\gamma}/R_{\rm BLR}$ in Figure 10, and we find most of the sources have $R_{\gamma}>R_{\rm BLR}$ with an average ratio of 10.

 \section{Conclusion and Discussion}  
    Using the simultaneous or quasi-simultaneous multi-wavelength observations from $Fermi$, $Swift$, $Planck$ and some ground-based telescopes, we model the SEDs of 25 blazars with a one-zone leptonic model, where the seed photons from BLR and molecular torus are considered in the EC process respectively. Due to rough agreement of the magnetic field strength in the  $\gamma$-ray emitting region and that of the radio core at pc scale\cite[e.g.,][]{from13,zama14,agar17}, we calculate the location of $\gamma$-ray emitting region for these blazars by assuming the magnetic field strength derived in the SED fitting follows the magnetic field strength distribution as derived from the radio core-shift measurements. The main results are summarized as follows, 1) we find that the SED fitting with the seed photons from the dusty torus is better than those from the BLR in LSP blazars, which supports the notion that the $\gamma$-ray emitting region may stays outside of the BLR; 2) we further calculate the location of the $\gamma$-ray emitting region for each selected blazar, and find that it locates at $\sim2\times10^{4}R_{\rm S}$ or $\sim 10 R_{\rm BLR}$ (see Table 2).  
 

  In our SED fitting, the sub-millimeter ($>200$ GHz) to $\gamma$-ray emission is considered, where the roughly simultaneous outbursts in these wavelengths suggest that they come from more or less a similar region \cite[e.g.,][]{siko08,wehr12,damm13,orie13,cohe14,li15,kush17}. It should be reasonable to model the multi-wavelength SEDs from sub-millimeter to $\gamma$-rays for these blazars with a one-zone model. In the LSP blazars, we find that the molecular torus seed photons in the EC process are better than those of BLR based on the MCMC analysis \cite[see also][]{cao13,kang14}. The reason for better SED fitting with torus seed photons maybe due to KN effect, where the high energy emission will be significantly suppressed for $\nu>10^{23}$ Hz for seed photons from BLR while this limit is $\sim4\times10^{25}$ Hz for torus seed photons \cite[see][for a more discussion]{kang14}. There is no strong cut-off in the $\gamma$-ray spectrum as observed by $Fermi$, which may suggest that the EC process is indeed seeded with photons from the torus \cite[see also][]{liu06,cao13,siko09}. 
  
  The possible source of seed photons in the EC process gives a rough estimation on the location of the $\gamma$-ray emitting region for LSP blazars (EC process may be weak or absent in BL Lac objects). With the MCMC method, we get a similar conclusion with that of \cite{kang14}. In this work, we further constrain the location of the $\gamma$-ray emitting region through comparison of the magnetic field strength as constrained from the SED fitting with the magnetic field strength distribution as derived from core-shift measurements. \cite{tave98} proposed that the magnetic field strength and Doppler factor can be derived from the multi-wavelength SEDs based on the SSC/EC model. The Synchrotron peak frequency is $\nu_{\rm s} \simeq 3.7 \times 10^{6}\gamma_{\rm b}^{2} B \delta/(1+z)$, where $\gamma_{\rm b} \simeq (3\nu_{\rm c}/4\nu_{\rm s})^{1/2}$ and $\gamma_{\rm b} \simeq (3\nu_{\rm ec}/4\nu_{\rm ext}\Gamma)^{1/2}$ for SSC and EC processes respectively ($\nu_{\rm c}$ is the SSC peak frequency, $\nu_{\rm ext}$ is the typical frequency of external seed photons, $\nu_{\rm ec}$ represents the EC peak frequency, all these frequencies are the observed frequencies). Another relation related to the magnetic field strength and Doppler factor is the ratio of the Synchrotron and Compton luminosity. In the SSC process, the ratio of the peak Compton $\gamma$-ray and Synchrotron emission is $L_{\rm C}/L_{\rm S}\propto L_{\rm S}/r_{\rm blob}^2\delta^4 B^2$. In the EC process, $L_{\rm C}/L_{\rm S}  \simeq U_{\rm ext}^{'}/U_{\rm B}^{'}$, where $U_{\rm ext}^{'} \simeq 3\times10^{-4}\Gamma^{2}$ erg cm$^{-3}$ is the external energy density of the torus seed photon field from molecular torus and $U_{\rm B}^{'} \simeq B^{2}/8\pi$ is the magnetic field energy density(the primed quantities are in the jet frame). Therefore, the magnetic field strength of blazars can be constrained from the ratio of the peak frequencies, $\nu_{\rm C}/\nu_{\rm S}$, and the ratio of the peak luminosities, $L_{\rm C}/L_{\rm S}$ based on the SSC or EC model, which do not very sensitive to other model parameters. The magnetic field strength is normally around 0.1-several Gauss in the $\gamma$-ray emitting region of the blazars \cite[see also,][]{zhang12,ghis10,pacc14}. In this work, we assumed homogeneous photon energy density for within BLR and torus, where the seed photon field may be anisotropic due to the possible anisotropic distribution of BLR/torus \citep[e.g., disk-like BLR,][]{cz16}. We find the magnetic field strength vary only by a factor of two even the energy density of seed photons vary by a factor of 10. Therefore, our main conclusion will not change even if there is some anisotropic distribution in BLR/torus. Beside the BLR/torus seed photons, there are also some other possibilities for the seed photons (e.g., accretion disk, CMB) in EC process \cite[e.g.,][]{pot13,ghis09}. The absorption effect should be very important if the seed photons predominantly come from the accretion disk, and the BLR would also absorb part of high-energy photons, which is not evident in observations \cite[e.g.,][]{liu06,bai09,pout10,ster11}. The CMB photons strongly evolve with the redshift, and the blazar spectra should also be strongly dependent on redshift if the CMB photons are dominant in the EC process. There is no strong observational support for this scenario. More simultaneous or quasi-simultaneous observations will also further help to build better multi-wavelength SEDs for blazars, and, in particular, the infrared data and $\gamma$-ray data at around MeV that are absent or quite limited now, are expected in the near future.  
  
   Assuming equipartition of magnetic/particle energy and that the synchrotron self-absorption is dominant in a conical jet, the position of the core should follow $R_{\rm core}\propto \nu^{-1/k_{\rm r}}$ with $k_{\rm r}\simeq1$  \cite[][]{koni81}. The distribution of the constrained $k_{\rm r}$ values roughly around 1 does support above assumptions \cite[e.g.,][]{osul09,soko11,push15,from13,agar17}. In case of a conical jet, the magnetic field strength and the electron number density are assumed to decrease with increasing distance from the central engine following power-law dependencies, where $B\propto R^{-1}$ and $n_{\rm e}\propto R^{-2}$ \cite[][]{blan79}. The VLBI observations can resolve the pc-scale radio core and the magnetic field strength distribution can be derived. The magnetic field strength derived from the SED fitting is more or less similar to that derived from the core-shift measurements, which suggests that the $\gamma$-ray emitting region should not be far from the radio core at $\sim$ several to several tens GHz. We note that the $k_{\rm r}$ value may not be exactly equal to 1 in different blazars and, therefore, magnetic field strength distribution also does not exactly follow $B\propto R^{-1}$. However, it will not affect our main conclusion on the location of the $\gamma$-ray emitting region since the magnetic field strength as derived from the SED fitting is quite close to that derived from the core-shift measurements (i.e., both are $\sim$ Gauss at pc-scale). Our results show that the $\gamma$-ray emitting region is at $\sim2\times10^4R_{\rm S}$ or $\sim10R_{\rm BLR}$ for the blazars in our sample. There is also some independent evidence supporting our conclusion.  Based on the radio core size and the time delay between the radio band and $\gamma$-ray band, \cite{kara17} found that the $\gamma$-ray active region is located at $1.9\pm1.1$ pc away from the jet base for PKS 1502+106, which is quite consistent with our estimates of $\sim2$ pc in this work. The correlation between the millimeter and the $\gamma$-ray light curves also indicates that the $\gamma$-ray emitting region should be located at $>$14 pc \cite[][]{agud11b} for OJ 287, which roughly corresponds to $\sim10^4 R_{\rm S}$ for the BH mass of $10^{10}\msun$ \cite[e.g.,][]{valt12}.

  \section*{Acknowledgements}
  We thank the anonymous referee for very valuable comments and constructive suggestions. We thank the Fermi-LAT Collaboration and the Planck Collaboration for data publication, and acknowledge all agencies and institutes contributing to the data publication. This research has made use of data partly obtained from literature. This work is supported by the NSFC (grants 11573009, 11622324), the CAS “Light of West China” Program (D. H. Yan). L. Chen is supported by the NSFC (grants 11233006 and U1431123) and the CAS grant (QYZDJSSW-SYS023).
 
\newpage

\begin{figure*}
\centering
\includegraphics[height=65mm]{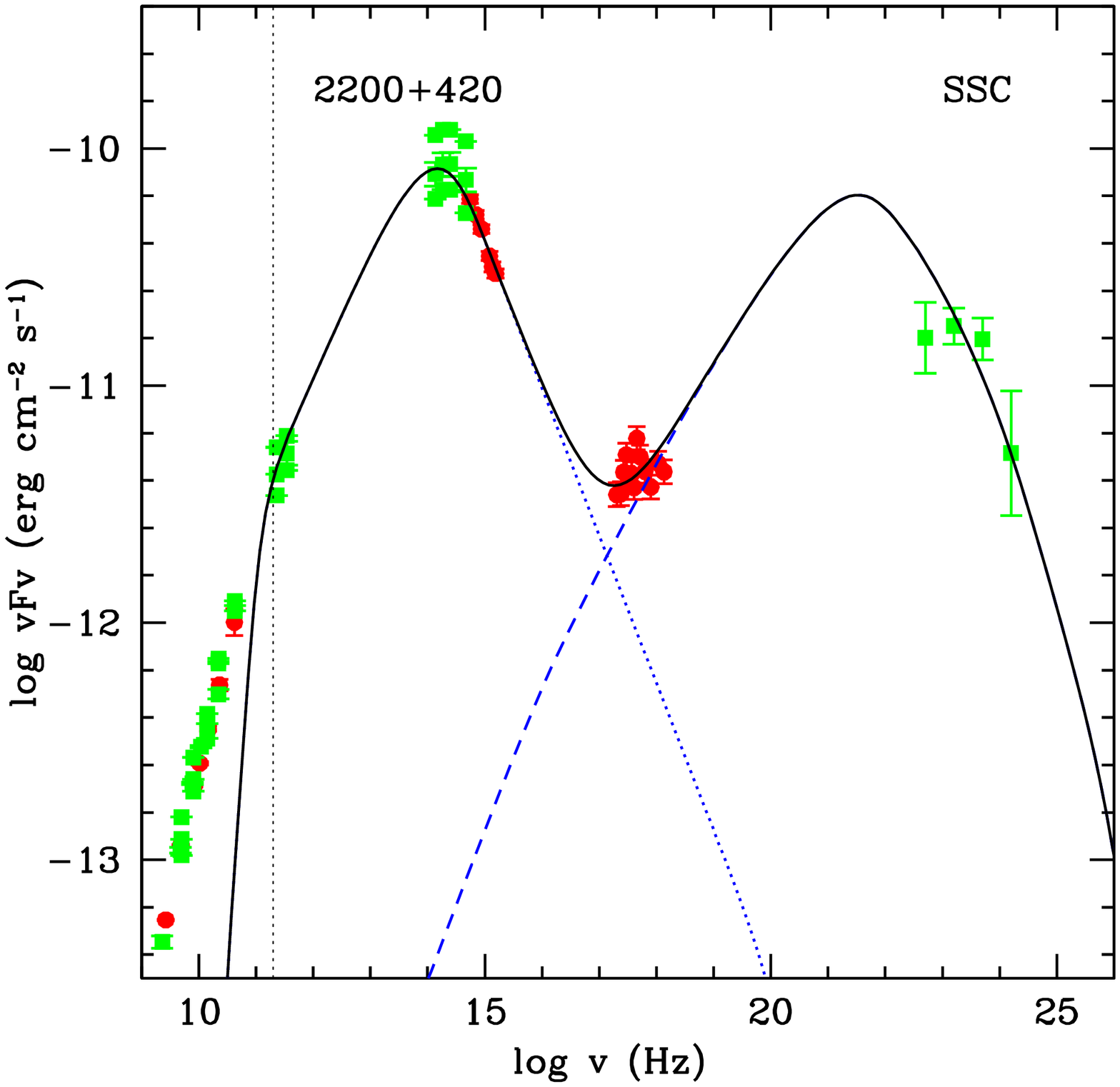}
\includegraphics[width=90mm]{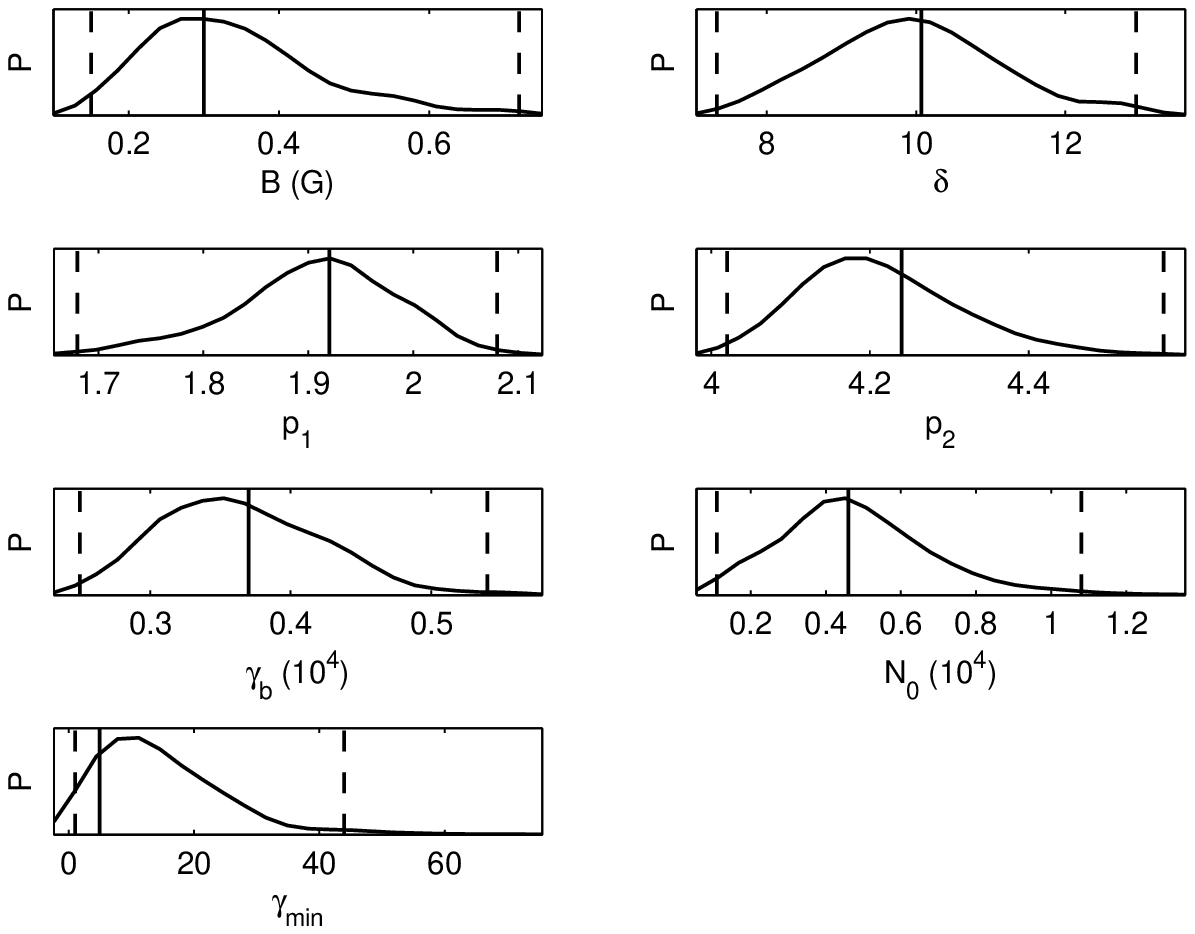}
\caption{ Left panel: the SED of 2200+420 (BL Lac) and its model, where only SSC is considered in the inverse Compton process in SED fitting and EC process is not important. The red solid circles represent simultaneous observational data, green solid squares present quasi-simultaneous data. The 200 GHz limit is shown with a vertical dotted line. The dotted, dashed and solid lines represent Synchrotron, SSC and total emission respectively. Right panel: one-dimensional parameters probability distributions for 2200+420, for each parameter the best fit value is presented by solid vertical line and the $68\%$ limits are presented by dashed vertical lines.}
\end{figure*}

\begin{figure*}
\includegraphics[width=190mm]{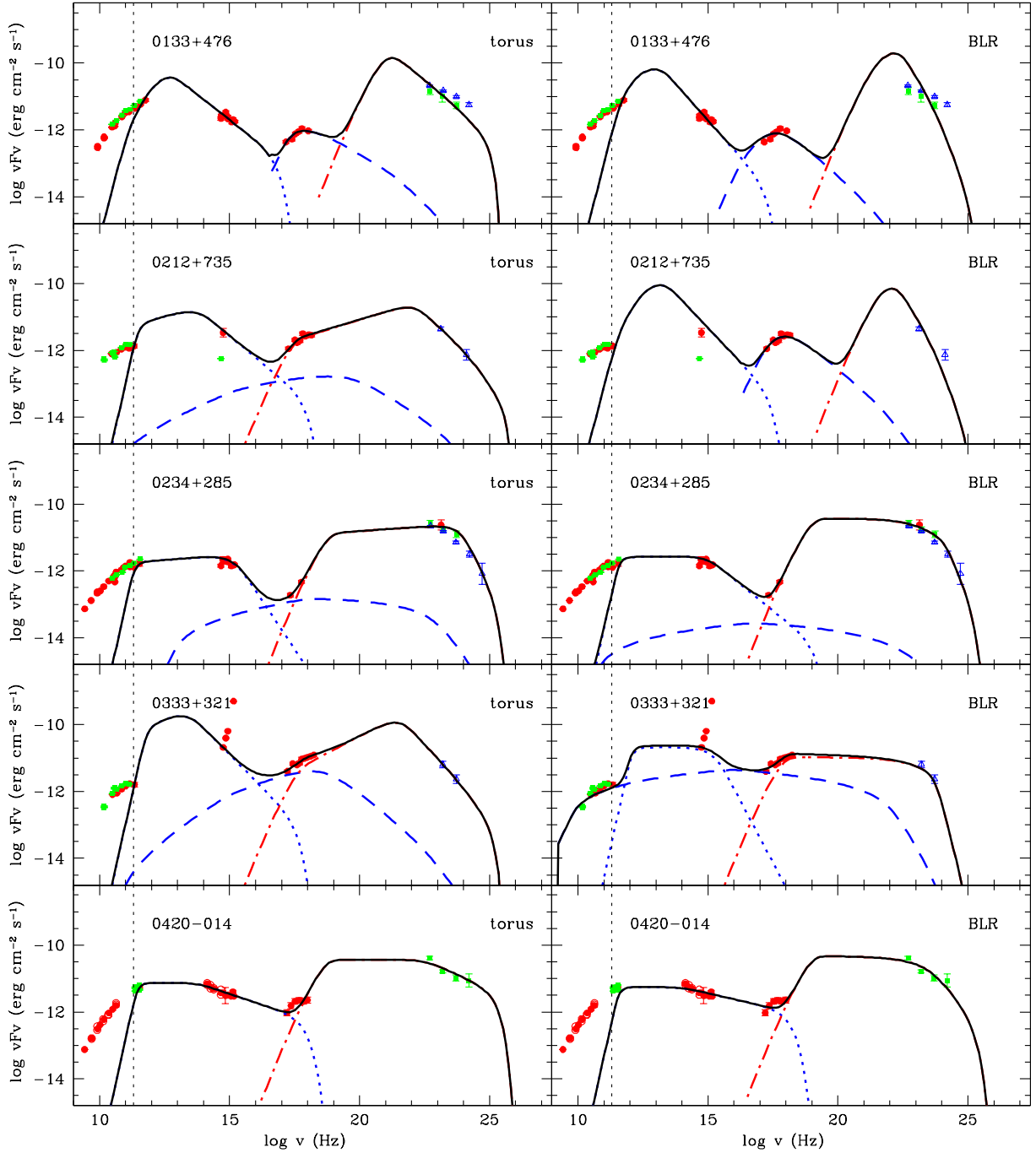}
\caption{The observed SEDs (data points) with model fitting (lines) for 0133+476, 0212+735, 0234+285, 0333+321 and 0420-014. Red filled circles represent simultaneous data, green filled squares represent quasi-simultaneous data, red open circles represent the simultaneous data observed at other epochs, and the blue triangles represent Fermi data integrated over 27 months. The 200 GHz limit is shown with vertical dotted lines. The left and right panels represent the fittings using seed photons originated from molecular torus and the BLR respectively in the EC process. The dotted, dashed, dot-dashed and solid lines represent synchrotron, SSC, EC and total emission respectively.}
\end{figure*}

\begin{figure*}
\centering
\includegraphics[height=200mm]{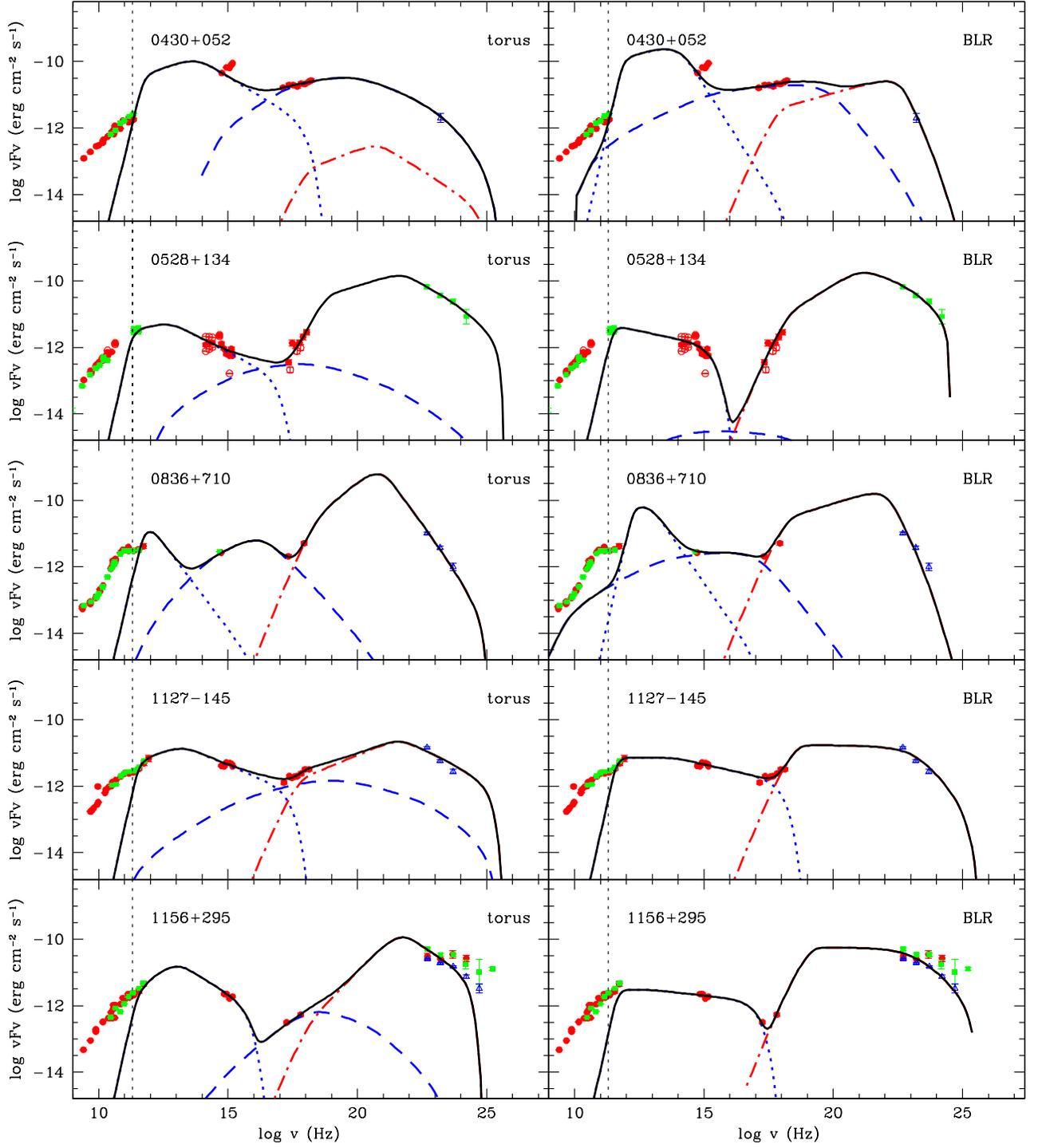}
\caption{The same as Figure 2, but for 0430+052, 0528+134, 0836+710, 1127-145 and 1156+295.}
\end{figure*}

\begin{figure*}
\centering
\includegraphics[height=200mm]{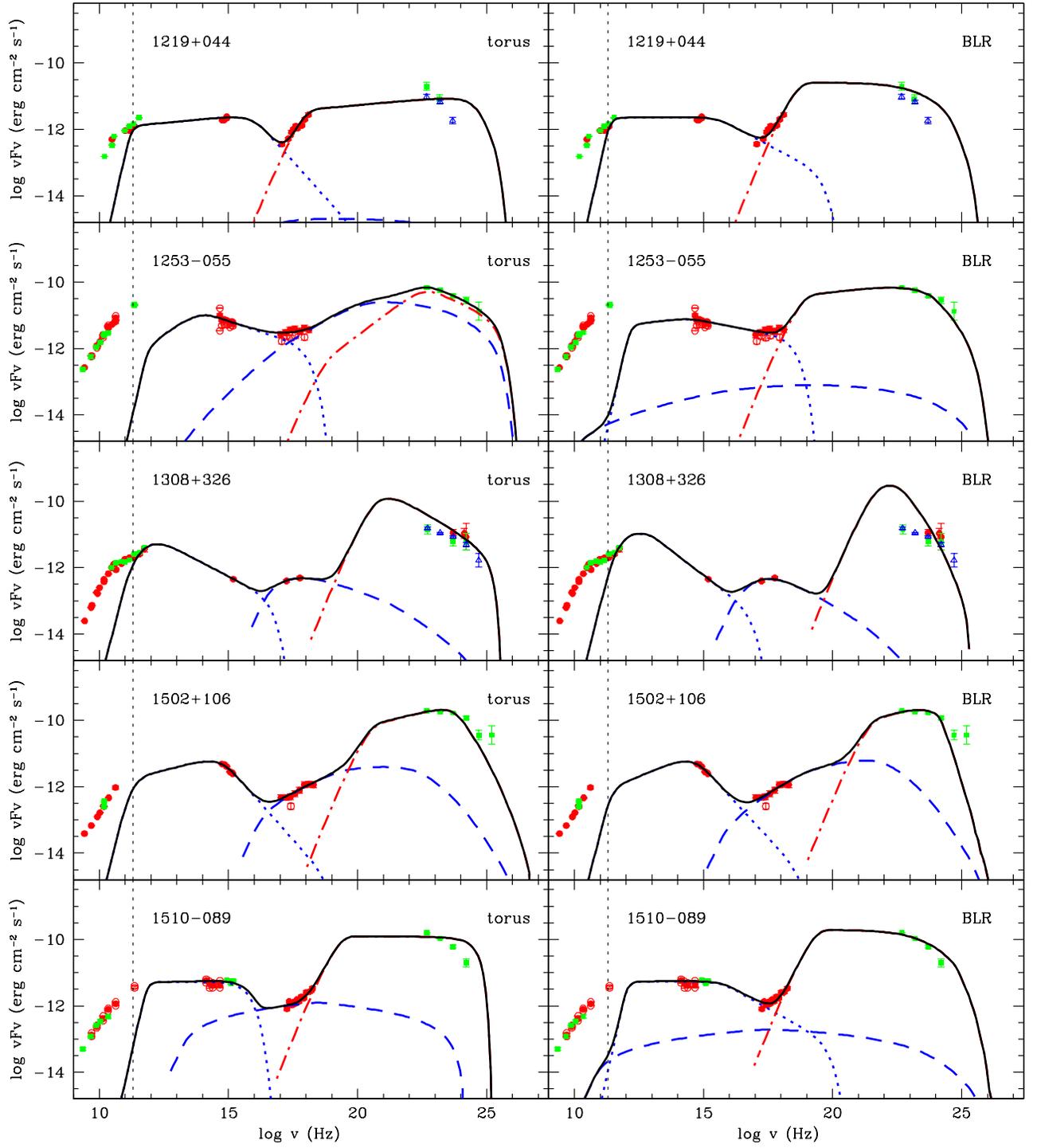}
\caption{The same as Figure 2, but for 1219+044, 1253-055, 1308+326, 1502+106 and 1510-089.}
\end{figure*}

\begin{figure*}
\centering
\includegraphics[height=200mm]{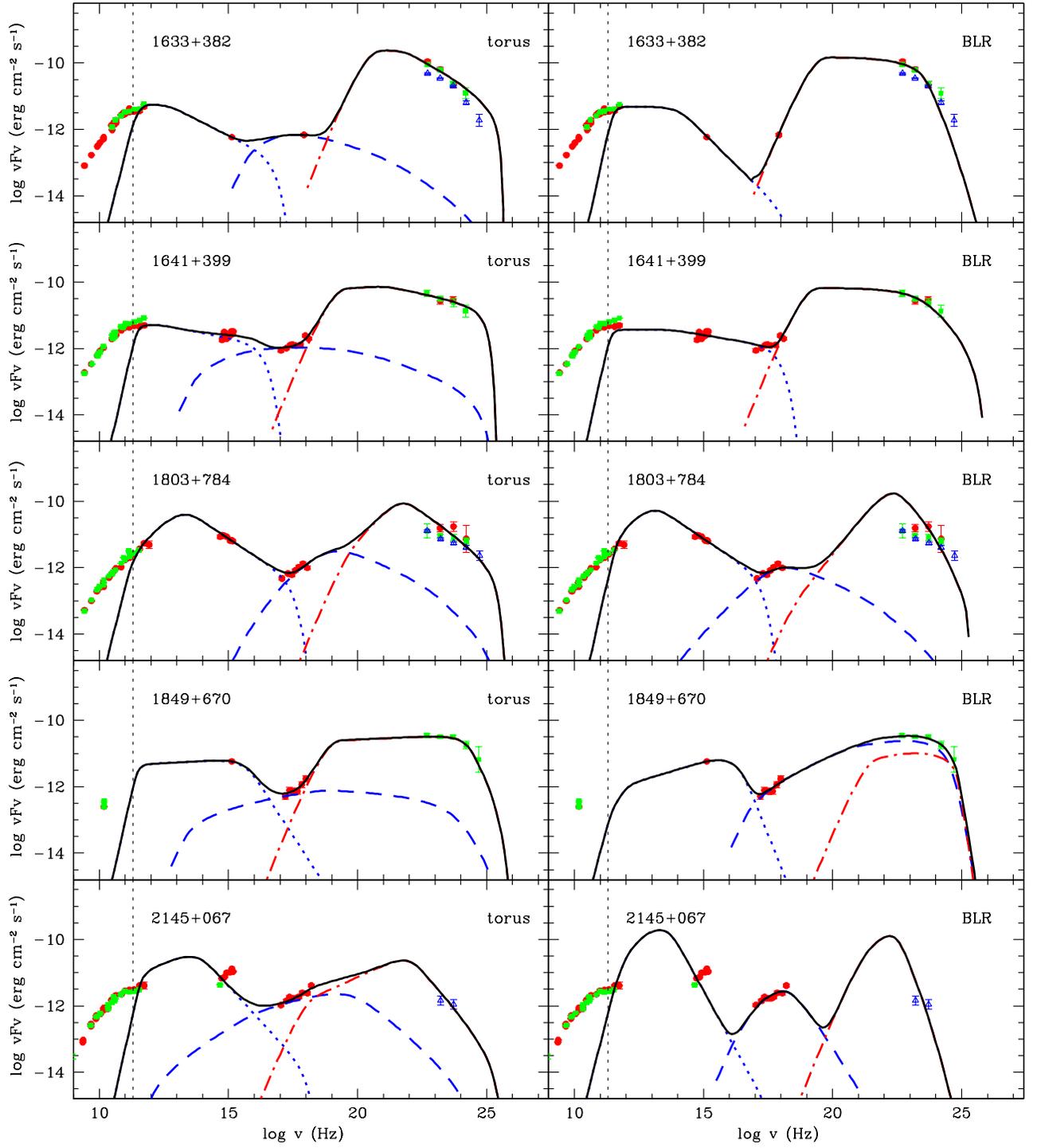}
\caption{The same as Figure 2, but for 1633+382, 1641+399, 1803+784, 1849+670 and 2145+067.}
\end{figure*}

\begin{figure*}
\centering
\includegraphics[height=200mm]{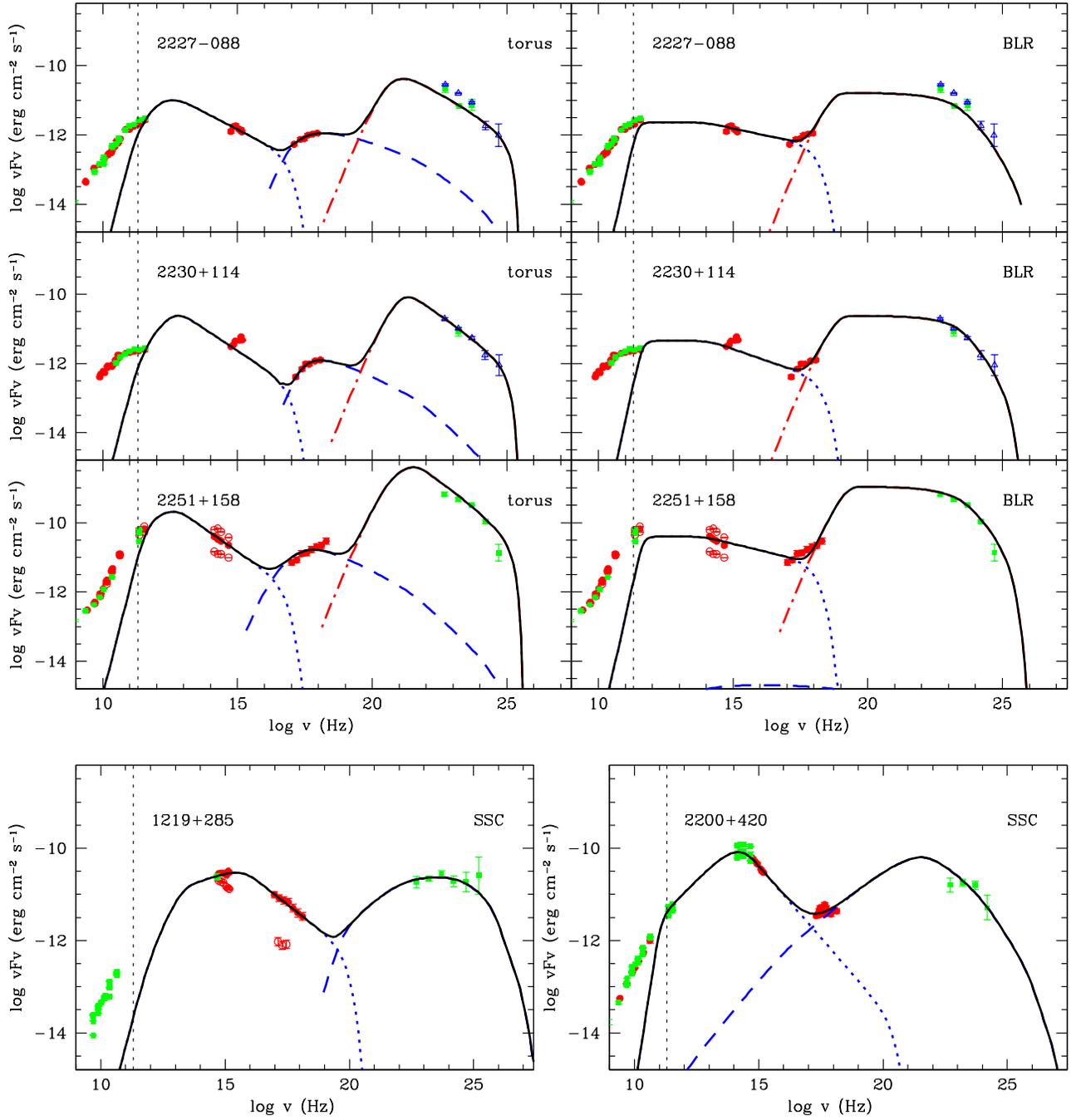}
\caption{The same as Figure 2, but for 2227-088, 2230+114, and 2251+158. For two BL Lacs (1219+285 and 2200+420), only Synchrotron and SSC are considered in SED \text{fitting}.}\end{figure*}

\begin{figure*}
\centering
\includegraphics[width=85mm]{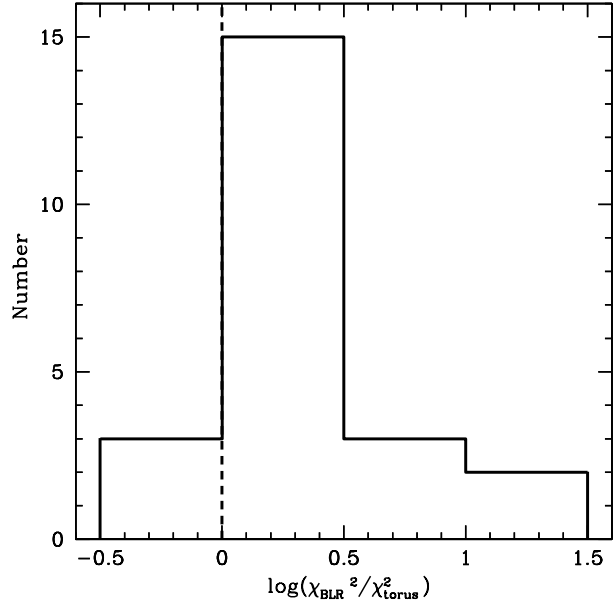}
\caption{The distribution of the $\log(\chi^{2}_{\rm BLR}/\chi^{2}_{\rm torus})$ for 23 LSP blazars, where $\chi^{2}_{\rm BLR}$ and $\chi^{2}_{\rm torus}$ represent the fitting result with seed photons from BLR and molecular torus respectively. The dashed line represents $\log(\chi^{2}_{\rm BLR}/\chi^{2}_{\rm torus})=0$}.
\end{figure*}

\begin{figure*}
\centering
\includegraphics[width=105mm]{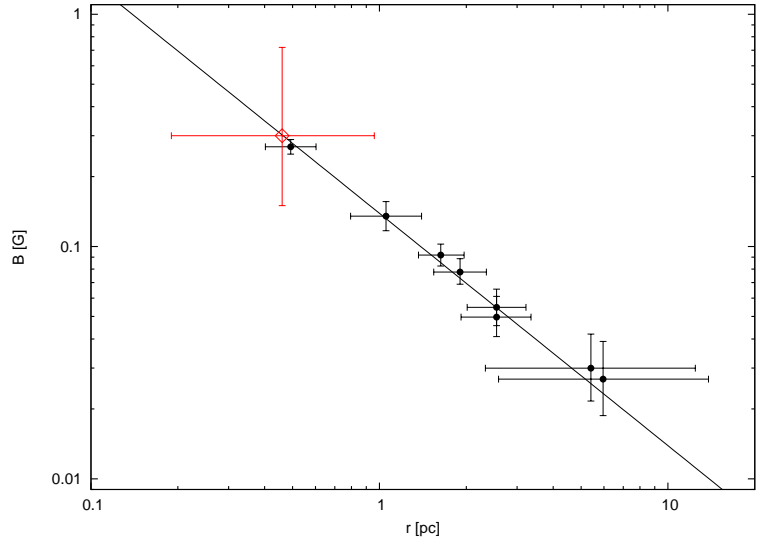}
\caption{The determination of the location of the $\gamma$-ray emitting region based on the magnetic field distribution of the radio core-shift measurements for 2200+420 (BL Lac), where core-shift measurements selected from \cite{osul09} are shown with black solid circles and the black line represents the best fit with a function of $B=B_1 R^{-1}$ ($B_1$ is the magnetic field strength at 1 pc).}
\end{figure*}

\begin{figure*}
\centering
\includegraphics[height=85mm]{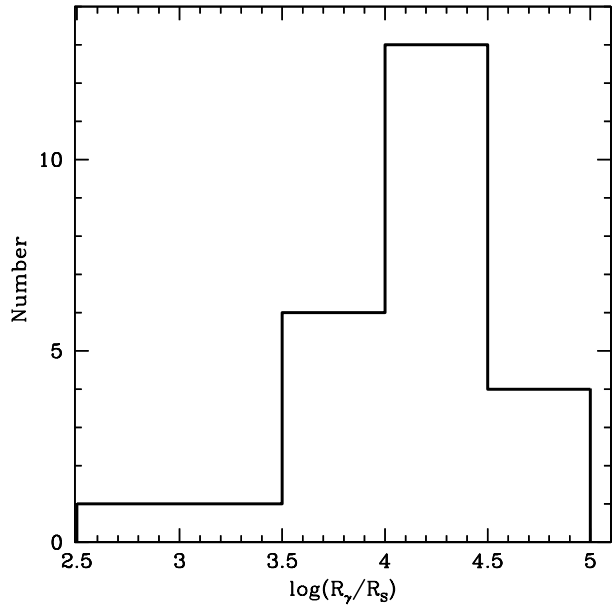} 
\caption{The distribution of the distance between the $\gamma$-ray emitting region and the BH ($R_{\gamma}$ \textbf{in units} of $R_{\rm S}$) and for 25 blazars in our sample.}
\end{figure*}

\begin{figure*}
\centering
\includegraphics[width=85mm]{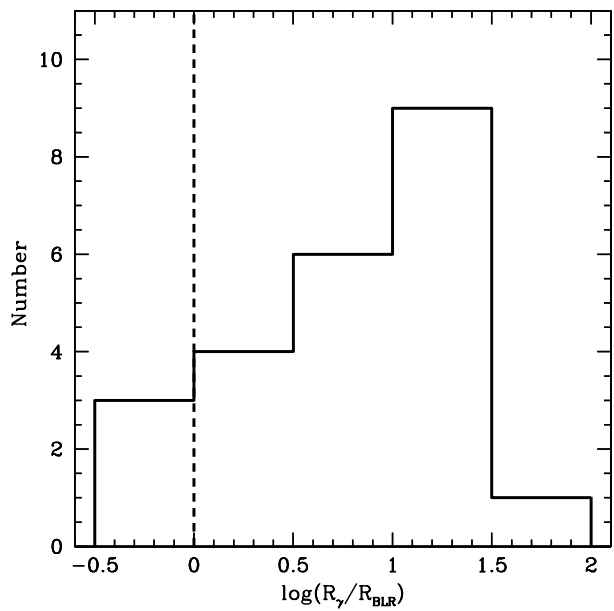}
\caption{The distribution of $R_{\gamma}/R_{\rm BLR}$ for 23 LSP blazars, where two BL Lacs without the putative BLR are not included (2200+420 and 1219+285).}
\end{figure*}


\begin{longrotatetable}
\begin{deluxetable*}{ccccccccccccccc}
\footnotesize
\tablecaption{SED fitting parameters}
\tablewidth{400pt}
\tabletypesize{\scriptsize}
\tablehead{\colhead{Object}&\colhead{Source name}&\colhead{Type}&\colhead{z$^{*}$}&\colhead{$t_{\rm var}(day)$}&\colhead{$B$(G)}&\colhead{$\delta$}&\colhead{$p_{1}$}& \colhead{$p_{2}$}&\colhead{${\gamma}_{\rm b}(10^{2})$}&\colhead{$N_{\rm 0}(10^{4})$}&\colhead{${\gamma}_{\rm min}$}&\colhead{$\frac{{\chi}_{\rm BLR}^{2}}{{\chi}_{\rm torus}^{2}}$}&\colhead{data$^{3}$}\\
(1) & (2) & (3) & (4) & (5) & (6) & (7) & (8) & (9) & (10) & (11) & (12) & (13) & (14) }

\startdata
\hline\hline
0133+476  &	S4 0133+47 & FSRQ &	0.86 &   ...  &$1.2_{-0.2}^{+0.2}$&$25.2_{-1.5}^{+1.5} $&$2.4_{-0.3}^{+0.3} $&$4.2_{-0.2}^{+0.1}$&$2.5_{-1.5}^{+0.5}$&$2.4_{-1.5}^{+10.7} $&$157_{-1}^{+19}$& 4.19   & 2B \\ 
0212+735  &1Jy 0212+735& FSRQ&2.37 &   ...  &$2.4_{-1.0}^{+2.0}$&$31.6_{-7.0}^{+10.1}$&$2.5_{-0.2}^{+0.4}$&$4.3_{-0.9}^{+2.7}$ &$7_{-5}^{+58}$&$0.5_{-0.4}^{+1.9}$&$3.2_{-0.9}^{+2.1}$&  7.34  & 2D \\
0234+285 &  4C 28.07   & FSRQ  &1.21 &   ...  &$0.7_{-0.1}^{+0.2}$&$23.3_{-2.7}^{+5.2}$&$2.9_{-0.1}^{+0.1}$&$5.3_{-1.3}^{+0.2}$ &$57_{-24}^{+63}$ &$14_{-11}^{+27}$ &$16_{-6}^{+8}$& 1.06   &	2C \\
0333+321 &	NRAO 140   & FSRQ&1.26&   ...   &$3.3_{-1.2}^{+1.2}$&$29.3_{-5.9}^{+3.6}$&$2.2_{-0.3}^{+0.6}$&$4.6_{-0.9}^{+1.5}$&$3_{-2}^{+9}$&$0.3_{-0.2}^{+1.8}$&$4_{-2}^{+3}$&1.88   &	2D  \\
0420-014 & PKS 0420-01 & FSRQ  &0.91&$0.517^{1}$&$2.3_{-0.6}^{+0.4}$&$58.4_{-7.2}^{+8.0}$&$2.99_{-0.22}^{+0.01}$&$3.5_{-0.1}^{+0.2}$&$4_{-1}^{+3}$&$0.05_{-0.04}^{+1.01}$&$5_{-2}^{+2}$& 2.02 	 &  1A  \\
0430+052 &   3C120    & FSRQ &0.03 &   ...	  &$6.0_{-3.4}^{+4.0}$&$3.5_{-0.6}^{+0.4}$&$2.4_{-0.1}^{+0.4}$&$3.9_{-0.4}^{+3.0}$&$8_{-4}^{+47}$&$55_{-17}^{+451}$&$33_{-33}^{+61}$&   1.25 &	2D  \\
0528+134 &PKS 0528+134 & FSRQ  &2.07&$1.05^{1}$&$0.5_{-0.1}^{+0.2}$&$31.5_{-2.9}^{+7.0}$&$2.5_{-0.2}^{+0.5}$&$3.8_{-0.2}^{+0.3}$&$5_{-2}^{+10}$&$3_{-1}^{+41}$&$13_{-2}^{+8}$& 0.95  &	1A  \\
0836+710 &	4C 71.07  &FSRQ   &2.22&$0.297^{2}$&$0.6_{-0.3}^{+1.7}$&$39.7_{-3.3}^{+9.3}$&$1.6_{-0.2}^{+0.6}$&$5.1_{-0.5}^{+0.4}$&$1.4_{-0.5}^{+0.5}$&$3_{-2}^{+17}$&$9_{-2}^{+4}$&  21.15 & 	2D  \\
1127-145 &PKS 1127-145&FSRQ   &1.18&   ...   &$1.4_{-0.2}^{+0.3}$&$19.7_{-1.6}^{+4.4}$&$2.4_{-0.2}^{+0.6}$&$3.7_{-0.3}^{+0.2}$&$6_{-3}^{+11}$&$3_{-2}^{+53}$&$6_{-2}^{+10}$&  0.92 &   2D  \\
1156+295 &  4C 29.45  &	FSRQ   &0.73&$0.29^{1}$&$1.0_{-0.1}^{+0.1} $&$30.6_{-2.3}^{+2.4} $&$1.4_{-0.2}^{+0.1} $&$4.0_{-0.1}^{+0.2} $&$3.4_{-0.3}^{+0.7}$ &$0.018_{-0.011}^ {+0.003}$&$6_{-6}^{+52}$& 1.44  &  2B  \\
1219+044 & PKS1219+04 &FSRQ  &0.96&   ...    &$1.4_{-0.5}^{+0.3}$&$32.6_{-6.8}^{+2.3}$ &$2.9_{-0.1}^{+0.1}$&$4.7_{-0.9}^{+0.8}$&$84_{-47}^{+68}$&$12_{-2}^{+132}$&$4.5_{-1.0}^{+3.0}$& 1.31  &	2C  \\
1219+285 &  W Coma & BL Lac&0.10&$0.044^{1}$&$0.10_{-0.04}^{+0.03}$&$42.8_{-3.7}^{+7.2}$&$2.63_{-0.03}^{+0.06}$&$3.9_{-0.2}^{+0.3}$&$180_{-50}^{+130}$&$330_{-8}^{+16}$&$890_{-20}^{+20}$& ...  &   1A  \\
1253-055 &  3C 279    &FSRQ  &0.54&$0.057^{1}$&$0.91_{-0.04}^{+6.72}$&$23.7_{-14.8}^{+1.4}$&$1.8_{-0.6}^{+0.1}$&$3.5_{-0.2}^{+0.1}$&$12_{-4}^{+5}$&$14_{-13}^{+15}$&$14_{-14}^{+106} $& 2.16   &	1A  \\
1308+326 &1Jy 1308+326&FSRQ  &	1.00  &   ...    &$0.4_{-0.1}^{+0.1}$&$22.6_{-2.6}^{+2.6}$&$2.97_{-0.33}^{+0.03}$&$3.8_{-0.3}^{+1.0}$&$4_{-3}^{+20}$&$140_{-100}^{+30}$&$140_{-60}^{+20}$ & 9.57  &	2B  \\
1502+106 &PKS 1502+106 & FSRQ &1.84&   ...   &$0.34_{-0.05}^{+0.05}$&$24.0_{-3.0}^{+2.9}$&$2.6_{-0.2}^{+0.2}$&$4.7_{-0.3}^{+0.6}$&$58_{-9}^{+12}$&$28_{-17}^{+53}$&$134_{-72}^{+82}$& 1.18  &	1A  \\
1510-089 &PKS 1510-08 &FSRQ  &0.36&$0.08^{2}$&$0.6_{-0.1}^{+0.1}$&$32.8_{-1.0}^{+1.0}$&$2.99_{-0.03}^{+0.01}$&$6.4_{-1.8}^{+0.6}$&$95_{-47}^{+100}$&$840_{-200}^{+160}$&$17_{-2}^{+2}$& 0.57  &	1A  \\
1633+382 & 4C 38.41& FSRQ  &1.81&$0.666^{1}$&$0.4_{-0.1}^{+0.3}$&$32.6_{-2.9}^{+2.3}$&$2.9_{-0.3}^{+0.1}$&$3.8_{-0.3}^{+1.8}$&$4_{-3}^{+23}$&$100_{-60}^{+80}$&$90_{-60}^{+40}$& 2.39  &	2B  \\
1641+399 & 3C 345  & FSRQ &0.59&	  ...    &$0.42_{-0.05}^{+0.05}$&$17.9_{-1.1}^{+1.0}$&$2.89_{-0.19}^{+0.05}$&$3.30_{-0.08}^{+0.10}$&$3_{-1}^{+2}$&$110_{-59}^{+6}$&$28_{-4}^{+7}$& 1.26  &	2B  \\
1803+784 &S5 1803+784 &BL Lac &0.68&   ...   &$1.2_{-0.3}^{+0.2}$&$19.6_{-1.1}^{+0.9}$&$1.20_{-0.02}^{+0.54}$&$4.0_{-0.1}^{+0.2}$&$5.4_{-0.3}^{+3.6}$&$0.003_{-0.001}^{+0.037}$&$52_{-1}^{+1}$& 1.78  &	2B \\
1849+670 &4C 66.20& FSRQ &0.66&   ...   &$0.7_{-0.2}^{+0.7}$&$18.7_{-12}^{+1.8}$&$2.92_{-0.64}^{+0.08}$&$5.4_{-1.3}^{+1.6}$&$100_{-30}^{+240}$&$40_{-10}^{+390}$&$20_{-5}^{+287}$  & 1.47 &   1A   \\
2145+067 & 4C 06.69 &FSRQ &0.99&   ...   &$2.0_{-0.6}^{+0.6}$&$19.2_{-1.6}^{+2.5}$&$2.3_{-0.1}^{+0.6}$&$4.7_{-0.8}^{+1.6}$&$8_{-4}^{+9}$&$1.5_{-0.5}^{+11.1}$&$8_{-5}^{+6}$  &	 3.02 &   2C   \\
2200+420 & BL Lac	 & BL Lac&0.07&$1.06^{2}$&$0.3_{-0.1}^{+0.4}$&$10.1_{-2.8}^{+2.9}$&$1.9_{-0.2}^{+0.2}$&$4.2_{-0.2}^{+0.3}$&$37_{-12}^{+17}$&$0.5_{-0.4}^{+0.6}$&$5_{-4}^{+39}$& ...    &   1A   \\
2227-088 & PKS 2227-08& FSRQ &1.56&   ...   &$1.0_{-0.2}^{+0.2}$&$22.2_{-2.4}^{+3.0}$&$2.7_{-0.1}^{+0.2}$&$3.8_{-0.3}^{+0.4}$&$4_{-2}^{+7}$&$27_{-11}^{+26}$&$147_{-61}^{+31}$ &  2.09 &	2C  \\
2230+114 & 4C 11.69  & FSRQ	 &1.04&   ...   &$1.4_{-0.2}^{+0.1}$&$28.1_{-2.2}^{+2.6}$&$2.3_{-0.5}^{+0.1}$&$4.0_{-0.2}^{+0.2}$&$2.5_{1.1}^{1.2}$&$4_{-4}^{+1}$&$155_{-26}^{+9}$  & 18.41 &	2D  \\
2251+158 &  3C 454.3 & FSRQ  &0.86&	  ...	&$0.7_{-0.1}^{+0.1}$&$34.0_{3.0}^{2.1}$&$1.30_{-0.07}^{+0.43}$&$4.1_{-0.2}^{+0.2}$&$2.5_{-0.5}^{+0.6}$&$0.01_{-0.0004}^{+0.08}$&$80_{-36}^{+5}$&  2.66  &	1A  \\
\enddata
\end{deluxetable*}
\tiny
\footnotesize
Notes.
Columns 1-3 show the name and source type. Columns 4 and 5 show the redshift and minimum variability timescale from literatures. Columns 6-12 are SED fitting parameters.
Column 13 is $\chi^{2}$ ratio of $\chi^{2}_{\rm BLR}$ and $\chi^{2}_{\rm torus}$. Column 14 is $\gamma$-ray data type.
1$\sigma$ errors of each SED fitting parameter are given in table.
 $\gamma_{\rm min}$ forced to be larger than 1.
$^{*}$ Redshifts given in \cite{zama14}.
$^{1}$The \textbf{minimum} optical variability timescale from \cite{liang03}.
$^{2}$The \textbf{minimum} $\gamma$-ray variability timescale from \cite{vovk13}.
$^{3}$The adopted $\gamma$-ray data in the fitting, where 1A represents the quasi-simultaneous data in \cite{abdo10a}, 2B represents the simultaneous data in \cite{giom12}, 2C represents the quasi-simultaneous data in \cite{giom12}, and 2D represents the integrated data over 27 months in \cite{giom12}.
\end{longrotatetable}


\newpage

\begin{table*}[ht]
\small
\caption{Location of $\gamma$-ray emission region and estimation of the  BLR size and the corresponding BH mass  }
\tabcolsep 1.2mm
\begin{tabular}{cccccccccccc}\hline\hline
Object    &$B_{\rm 1pc}$$^{a}$&$B$&$R_{\rm \gamma}$& line type &     FWHM     &  $L_{\rm line}$  & reference & $R_{\rm BLR}$& $\log {M}$  \\
          &     (G)        &     (G)       &  $(10^{17}cm)$  & & ( km s$^{-1}$) &$(10^{42}\ergs)$ &           &  $(10^{17}cm)$ &  $(\msun)$\\
    (1) & (2) & (3) & (4) & (5) & (6) & (7) & (8) & (9) & (10) \\
\hline\hline
0133+476   &	0.77 &$1.2_{-0.2}^{+0.2}$ &$20_{-3}^{+4}$  &	H$\beta$ &4223 &20.97& Tor12 & 5.0 & 8.7  \\
0212+735   &	1.27 &$2.4_{-1.0}^{+2.0}$&$16_{-8}^{+12}$&  C IV  & 5579 & 11345.32 & Tor12 & 34.4& 9.6  \\
0234+285   &	1.71 &$0.7_{-0.1}^{+0.2}$& $77_{-17}^{+16}$&  Mg II & 5100 & 148.59 & Shaw12& 6.1 & 9.1  \\
0333+321   &	1.95 &$3.3_{-1.2}^{+1.2}$&$18_{-5}^{+10}$&  Mg II & 2900 & 520.57   & Liu06 & 12.4& 9.0 \\
0420-014   &	1.41 &$2.3_{-0.6}^{+0.4}$&$19_{-3}^{+8}$&  Mg II & 4846 & 43.37   & Tor12 &	3.0 & 8.8  \\
0430+052   &	0.11 &$6.0_{-3.4}^{+4.0}$&$0.6_{-0.3}^{+0.7}$& H$\beta$  & 2750&  1.29   & Tor12 &	0.8 & 7.5  \\ 	
0528+134   &	1.60 &$0.5_{-0.1}^{+0.2}$&$96_{-24}^{+16}$&  C IV	 & ... &  2500   & Pal11 & 13.9&  9.0$^{c}$  \\
0836+710   &	1.93 &$0.6_{-0.3}^{+1.7}$&$106_{-80}^{+71}$&  C IV &7657& 1111.15  & Tor12 &	8.5 & 9.3 \\
1127-145   &    0.84 &$1.4_{-0.2}^{+0.3}$&$19_{-5}^{+3}$&  Mg II  & 5101 &	194.05 & Tor12 &7.0 & 9.2  \\
1156+295   &	1.17 &$1.0_{-0.1}^{+0.1} $&$36_{-4}^{+4}$&  Mg II	& 4245 &	28.77   & Tor12 &  2.4 & 8.6  \\
1219+044   &	0.81 &$1.4_{-0.5}^{+0.3}$&$18_{-4}^{+9}$&  Mg II	 & 5268 &	39.40  & Tor12 &	2.8 & 8.8  \\
1219+285   &	0.08 &$0.10_{-0.04}^{+0.03}$&$25_{-6}^{+18}$& ...  &  ...	   &   ...   &	...  &  ...  & 8.7$^{c}$   \\
1253-055   &$<0.42 $&$0.91_{-0.04}^{+6.72}$&$<14_{-13}^{+1}$& H$\beta$& 3100 &	17.28  & Liu06 & 4.4 & 8.4  \\
1308+326   &	0.96 &$0.4_{-0.1}^{+0.1}$&$72_{-13}^{+18}$&  Mg II	 & 5267 &	21.33  & Tor12 &	2.0 & 8.7  \\
1502+106   &	0.69 &$0.34_{-0.05}^{+0.05}$&$63_{-8}^{+10}$&  Mg II &  5000 & 66.22  & Shaw12& 3.8 & 8.9  \\
1510-089   & 0.73 &$0.6_{-0.1}^{+0.1}$&$38_{-3}^{+3}$& H$\beta$& 3250 &	21.94  & Tor12 & 5.2 & 8.5  \\
1633+382   &	1.62 &$0.4_{-0.1}^{+0.3}$&$115_{-38}^{+65}$&  Mg II  &5583&	78.33  & Tor12 &	4.2 & 9.0  \\
1641+399   &	1.20 &$0.42_{-0.05}^{+0.05}$&$88_{-10}^{+12}$&  Mg II & 5520 & 102.18  & Tor12 & 4.9 & 9.1  \\
1803+784   &$<0.39$&$1.2_{-0.3}^{+0.2}$&$<10_{-1}^{+3}$&	H$\beta$ & 4320  &  15.59  & Tor12 &	4.1 & 8.6 \\
1849+670   &	0.52 &$0.7_{-0.2}^{+0.7}$&$24_{-12}^{+13}$&  Mg II	 & 5868 &	20.85  & Tor12 &	2.0 & 8.8 \\
2145+067   &$<0.34$&$2.0_{-0.6}^{+0.6}$&$ <5_{-1}^{+2} $&  Mg II	 & 5517	 &  457.53 & Tor12 & 11.5 & 9.5 \\
2200+420   &$0.139^{b}$&$0.3_{-0.1}^{+0.4}$&$14_{-8}^{+15}$& ... &   ...   &   ...   &  ...  &  ...  & 8.2$^{c}$   \\
2227-088   &	1.44 &$1.0_{-0.2}^{+0.2}$&$45_{-7}^{+8}$&  Mg II	 & 5896 &	38.15  & Tor12 &	2.8 & 8.9  \\
2230+114   &	2.12 &$1.4_{-0.2}^{+0.1}$&$48_{-5}^{+6}$&  Mg II	 & 4583 &	71.21  & Tor12 &4.0 & 8.9 \\
2251+158   &	0.32 &$0.7_{-0.1}^{+0.1}$&$14_{-1}^{+2}$&  Mg II	 & 5162 &	125.98 & Tor12 &	5.5 & 9.1 \\ 	
\hline\hline
\end{tabular}
\scriptsize
References:Tor12: \cite{tor12}; Shaw12: \cite{shaw12}; Liu06:\cite{liu06b}; Pal11:\cite{pal11}.\\
Notes.
Column 1 is source name. Columns 2 and 3 represent the magnetic field strength derived from core-shift measurements and SED fitting respectively. Column 4 is location of the $\gamma$-ray emission region. Columns 5-8 represent broad-emission-line types, FWHM, luminosities and related references. Columns 9 and 10 are BLR size and the BH mass.\\
$^{a}$ $B_{\rm 1pc}$ selected from \cite{zama14}.
$^{b}$ $B_{\rm 1pc}$ adopted from \cite{osul09}.
$^{c}$ BH mass obtained from \cite{zama14}.
\end{table*}

\end{document}